\documentclass[aps,prb,reprint,twocolumn,superscriptaddress,showpacs,floatfix]{revtex4-1}
\usepackage{hyperref}
\usepackage{graphicx}
\usepackage{amsmath}
\usepackage{amssymb}
\usepackage{enumitem}
\begin{document}
\title{Combining Density Functional Theory and Green's Function Theory: Range-Separated, Non-local,
Dynamic, and Orbital-Dependent Hybrid Functional} 
\author{Alexei A. Kananenka}
\affiliation{Department of Chemistry, University of Michigan, Ann Arbor,
Michigan 48109, United States}
\email{akanane@umich.edu}
\author{Dominika Zgid}
\affiliation{Department of Chemistry, University of Michigan, Ann Arbor,
Michigan  48109, United States}

\begin{abstract}

We present a rigorous framework that combines single-particle Green's function theory with density functional
theory based on a separation of electron-electron interactions into short-range and long-range
components. Short-range contributions to the total energy and exchange-correlation 
potential
are provided by a density functional approximation, while the long-range contribution is
calculated using an explicit many-body Green's function method. Such a hybrid  results in 
a nonlocal, dynamic, and orbital-dependent exchange-correlation functional of a single-particle Green's function.
In particular, we present a range-separated hybrid functional called srSVWN5---lrGF2 which 
combines the  local-density approximation and the second-order Green's function theory. We illustrate 
that similarly to density functional 
approximations the new functional is weakly basis-set dependent. Furthermore, it offers an improved 
description of the short-range dynamical correlation. The many-body contribution to the functional 
allows us to mitigate the many-electron self-interaction error present in most of density functional
approximations and provides a better description of molecular properties. 
Additionally, the new functional can be used to scale down the self-energy and, therefore, introduce an
additional sparsity to the self-energy matrix that in the future can be exploited in 
calculations for large molecules or periodic systems.
\end{abstract}

\maketitle


\section{Introduction}
\label{sec:intro}

Kohn--Sham density functional theory
(DFT)~\cite{Hohenberg:pr/136/B864,Kohn:pr/140/A1133,engel2011density} 
has become a method of choice for unraveling the ground state properties of 
mostly single reference molecular and condensed matter systems. 
Its popularity is due to an attractive compromise between the accuracy and 
computational cost, provided by numerous approximations to the exchange-correlation functional. 
The best approximate functionals offer a decent description of  the short-range dynamical correlation
which justifies their use for near-equilibrium geometries. Another attractive feature of 
density functionals is their weak dependence on the one-electron basis set. Despite their 
large success, however, local and semilocal density
functionals fail to describe a number of important properties, for example, charge
transfer excitations~\cite{Dreuw:jcp/119/2943}, dynamical long-range correlations important in weak 
van der Waals complexes bound by London dispersion forces~\cite{Kristyan:cpl/229/175}, and Rydberg 
excitation energies~\cite{Tozer:jcp/109/10180}. The reason for this failure is well understood and is 
rooting in a wrong asymptotic behavior of the  exchange-correlation potential which in turn is a consequence of 
a self-interaction error~\cite{Perdew:prb/23/5048}.

Many-body  wave-function methods such as the M{\o}ller--Plesset
perturbation theory (MP2)~\cite{Moller:pr/46/618}, coupled cluster (CC)~\cite{Bartlett:rmp/79/291} or 
multiconfigurational
self-consistent field (MCSCF)~\cite{roos2016multiconfigurational} 
approaches are
capable of providing a correct description when density functionals fail. However, for these \textit{ab-initio} 
methods, in addition to a steep computational cost and long configuration expansion 
of the wave function also large  basis sets are required to describe
the dynamical correlation accurately and reach an agreement with experiments. 
These features make the application of  \textit{ab-initio} methods to very large systems quite 
challenging and much larger system sizes can be reached when density functional approximations are used.

In recent years, there has been a substantial progress in the development of density functionals that
mix both the standard local or semilocal density functional approximation with the wave-function theory.
The mixing is done rigorously by separating the two-electron interaction operator into 
short-range and long-range 
components~\cite{savinRSoriginal,Leininger:cpl/275/151,Toulouse:pra/70/062505} resulting in so-called 
range-separated hybrid functionals~\cite{Baer:arpc/61/85}.
They are meant to combine the best features of the respective approaches. The least computationally
expensive range-separated hybrid functional is obtained when a non-local 
Hartree--Fock-type exchange is introduced to replace the long-range exchange density 
functional~\cite{Yanai:cpl/393/51,Heyd:jcp/118/8207,Iikura:jcp/115/3450,Chai:jcp/128/084106,Vydrov:jcp/125/234109}. Such functionals were proved successful in a partial correction of the long-range behavior 
of the exchange-correlation potential~\cite{Iikura:jcp/115/3450,Tawada:jcp/120/8425}.
However, they are also known to perform worse than standard density functionals in some
cases~\cite{Yanai:cpl/393/51,Peach:pccp/8/558}.

The combination of explicit many-body wave-function methods with the density-functional theory 
by means of range separation has been previously quite extensively explored. 
Long-range 
MP2~\cite{Angyan:pra/72/012510,Fromager:pra/78/022504,Kullie:cp/395/54,Angyan:pra/78/022510}, 
second-order $n$-electron valence state perturbation theory (NEVPT2)~\cite{Fromager:pra/81/024502}, coupled cluster (CCSD(T))~\cite{Goll:pccp/7/3917}, 
random-phase approximation (RPA)~\cite{Janesko:jcp/130/081105,Toulouse:prl/102/096404,Toulouse:pra/82/032502}, configuration interaction (CI)~\cite{Leininger:cpl/275/151,Pollet:jcp/116/1250}, 
MCSCF~\cite{Fromager:jcp/131/054107,Fromager:jcp/126/074111}, and 
the density-matrix-functional theory~\cite{Pernal:pra/81/052511,Rohr:pra/82/052502} 
have been combined
with short-range local and semilocal density functionals~\cite{Toulouse:pra/70/062505,Goll:pccp/7/3917,Toulouse:ijqc/100/1047,Heyd:jcp/118/8207,GoriGiorgi:pra/73/032506}.
These range-separated functionals were successfully applied to weakly interacting molecular
systems~\cite{Gerber:cpl/416/370,Toulouse:pra/82/032502,Angyan:pra/72/012510,Goll:pccp/7/3917,Toulouse:prl/102/096404,Toulouse:pra/82/032502,Liu:jctc/7/2399,Zhu:jcp/132/244108,Paier:jcp/132/094103}. 
In comparison to corresponding standard many-body wave-function approaches, the range-separated functionals 
have additional advantages such as a rapid  convergence with respect to the basis set size~\cite{Odile:jcp/142/074107,Goll:pccp/7/3917,Angyan:pra/72/012510,Janesko:pccp/11/9677,Kullie:cp/395/54,Cornaton:pra/88/022516,Toulouse:prl/102/096404,Janesko:jcp/130/081105,Janesko:jcp/131/034110,Irelan:jcp/135/094105,Toulouse:pra/82/032502} and smaller basis-set superposition errors.
In these approaches, the long-range correlation energy is usually added as a post-SCF correction to the 
total energy from a range-separated calculation without the long-range correlation functional. Therefore,
they do not yield the exact energy even with the exact short-range exchange-correlation functional, 
for example see ref~\citenum{Toulouse:pra/82/032502}.

Since the srSVWN5---lrGF2 functional introduced in this work combines both the density functional theory and 
the Green's function theory, we aim to provide a self-contained and detailed description that can be
useful to both these communities. Therefore, to bring the readers to a common ground, we found it
helpful to list some key theory concepts from both communities.

Finite-temperature single-particle Green's function methods have been long known in the context of
condensed matter physics~\cite{fetter2003quantum,stefanucci2013nonequilibrium,abrikosov2012methods} and 
now are making inroads into quantum chemistry~\cite{Phillips:jcp/140/241101,Kananenka:prb/91/121111,Lan:jcp/143/241102,Lan:jpcl/8/2200,Phillips:jcp/142/194108}. 
%
%
These methods are rigorous and offer several advantages. 
The single-particle Green's function formalism is based entirely on 
one-electron operators avoiding the necessity of dealing with wave functions.  
A single-particle Green's function determines the expectation 
value of single-particle operators,
the two-electron correlation energy, and provides access to the spectral density, ionization potentials and 
electron affinities.

In this work, we present a rigorous self-consistent framework combining a short-range
density functional approximation with a long-range single-particle Green's function method.
As a specific example, we implemented and benchmarked the short-range local density
approximation (LDA)~\cite{Dirac:mpcps/26/376,Vosko:cjp:58/1200} with the second-order Green's function theory
(GF2)~\cite{Holleboom:jcp/93/5826,Phillips:jcp/140/241101,Dahlen:jcp/122/164102}.
%
%
To further motivate this work, it is worth to briefly list differences between the 
method presented here and the already existing plethora of range-separated hybrid functionals. 
Most methods that have been previously applied to the long-range interactions were not 
self-consistent. 
In contrast to non-self-consistent methods, which are starting point dependent,  the approach presented here, irrespective of the initial guess,  recovers 
the exact total electronic energy  provided that both the exact short-range 
exchange-correlation functional and the exact long-range Green's function method are used.
An iterative nature of GF2
results in multiple implications. 
The overall accuracy of GF2 for weakly correlated systems
is close to that of MP2 or CCSD, however, unlike these two approaches, GF2 does not display divergences for 
strongly correlated systems~\cite{Phillips:jcp/140/241101}. GF2 is a one-electron self-interaction 
free method, while methods such as RPA contain a significant one-electron self-interaction error~\cite{MoriSanchez:pra/85/042507}. Furthermore, a Matsubara axis GF2 formalism is explicitly temperature-dependent.


The range-separated hybrid functional presented here also shares some commonalities with other combinations of 
DFT with Green's function methods.  For example, the 
LDA+DMFT~\cite{Kotliar:rmp/78/865} method that combines LDA with the dynamical mean-field theory
(DMFT)~\cite{Georges:prb/45/6479} is often used in solid state calculations 
of strongly correlated systems. However, LDA+DMFT is known to suffer from a so-called 
double counting  problem~\cite{Kotliar:rmp/78/865,Lee:prb/91/155144}, where some 
electronic correlations 
are accounted for by both LDA and DMFT. In the LDA+DMFT method, these two sources of electronic 
correlations cannot be rigorously separated~\cite{Anisimov:jpcm/9/767,Haule:prb/81/195107,Kotliar:rmp/78/865}. 
We would like to stress that the double counting problem does not
appear in the framework presented here since the exact separation of the electron-electron 
interaction into long- and short-range components is used.

Range-separated hybrid functionals employ a single range separation parameter controlling the spatial
extent of the short-range contribution.
The optimal value of this system-dependent parameter~\cite{Rohrdanz:jcp/129/034107,Stein:prl/105/266802,Korzdorfer:jcp/135/204107} 
can be determined either by empirical fitting against available experimental
data~\cite{Chai:jcp/128/084106,Yanai:cpl/393/51,Livshits:pccp/9/2932,Song:jcp/126/154105} 
or in an \textit{ab-initio} fashion in a self-consistent procedure~\cite{Livshits:pccp/9/2932,Baer:arpc/61/85}. 
In our current work, we have adopted the latter view and applied the
optimal tuning strategy based on calculations of ionization potentials to find optimal values of 
the range separation parameter for several atoms and molecules. 
Additionally, we have also investigated the two-electron self-interaction error,
basis set dependence, dynamical correlation as well as the implications of the hybrid functional presented
here for the Green's function based embedding methods and periodic calculations.

\section{Theory}
\label{sec:theory}
The exact electronic ground state energy of a system of $N$ interacting electrons in the presence of 
external potential $v(\mathbf{r})$ (e.g., the potential of the nuclei) can be obtained by a two-step 
minimization of the following functional~\cite{Levy:pnas/76/6062} 
\begin{equation}
E_\text{tot}[\rho] = \underset{\rho \to N}{\min} \Bigl\{ 
F[\rho] + \int d\mathbf{r}v(\mathbf{r})\rho(\mathbf{r}) 
\Bigr\},   \label{eq:es}
\end{equation}
where  $\rho(\mathbf{r})$ is an electron density and $F[\rho]$ is the universal functional of the electron density that is defined as
\begin{equation}
F[\rho] = \underset{\Psi \to \rho}{\min} \langle \Psi | \hat{T} + \hat{V}_{ee} | \Psi \rangle,     \label{eq:hk_func}
\end{equation}
where $\hat{T}=-\frac{1}{2}\sum_i^N \nabla_i^2$ is the kinetic energy operator, $\hat{V}_{ee}=\frac{1}{2}\sum_{i\neq j}^N \hat{v}_{ee}(r_{ij})$ 
is the
electron-electron interaction operator, $r_{ij}=|\mathbf{r}_i - \mathbf{r}_j|$  and $\mathbf{r}_i$ is the
coordinate vector of electron $i$. The minimization is first carried out over all normalized antisymmetric 
wave functions $\Psi$ that produce a given density $\rho(\mathbf{r})$, and then over all densities
yielding $N$-electrons. The existence and uniqueness of the universal functional $F[\rho]$ is guaranteed by 
Hohenberg and Kohn theorem~\cite{Hohenberg:pr/136/B864}. 
Regrettably, an explicit variation of eq~\ref{eq:es}
has not become practical since no exact form of the universal functional is available and due to its 
absence all practical applications are based on the Kohn--Sham
scheme~\cite{Kohn:pr/140/A1133}. This procedure uses an approximation to the exchange-correlation 
part of the universal functional. One of the most successful approaches taken 
along this way is the combination of two (or more) density functional approximations into one so-called
hybrid exchange-correlation functional using the adiabatic connection
theorem~\cite{Harris:jpfmp:4/1170,Gunnarsson:prb/13/4274,Langreth:prb/15/2884,Harris:pra/29/1648}.

Range-separated density functional approximations belong to a particular class of hybrid 
functionals~\cite{Baer:arpc/61/85}.
The essence of range-separated hybrid functionals lies in the decomposition of the Coulomb 
electron-electron interaction operator into a sum of short-range and long-range 
counterparts~\cite{savinRSoriginal,Leininger:cpl/275/151,Dombroski:jcp/100/6272},
\begin{eqnarray}
\frac{1}{r_{ij}} = &\hat{v}_{ee}^{sr,\lambda}(r_{ij}) + \hat{v}_{ee}^{lr,\lambda}(r_{ij})\\ \nonumber
  &=
\underbrace{\frac{1-f\left(\lambda r_{ij}\right)}{r_{ij}}}_\text{short-range}+
\underbrace{\frac{f\left(\lambda r_{ij}\right)}{r_{ij}}}_\text{long-range},    \label{eq:decomp}
\end{eqnarray}
with the parameter $\lambda$ controlling the range separation. 
The function $f(\lambda r)$ satisfies the following properties 
$f(\lambda r \to \infty)=1$ and $f(\lambda r \to 0) = 0$. From a physical and computational
standpoint the standard error function $f(\lambda r)=\text{erf}(\lambda r)$ is one of the
most convenient choices. 
The decomposition in eq~\ref{eq:decomp} is exact and presents a convenient starting point for developing 
range-separated hybrid functionals by mixing a short-range density functional approximation with a 
long-range method. The universal functional from eq~\ref{eq:hk_func} is partitioned
accordingly~\cite{Toulouse:pra/70/062505}
\begin{equation}
F[\rho] = \underset{\Psi \to \rho}{\min} \langle \Psi^\lambda | \hat{T} + \hat{V}^{lr,\lambda}_{ee} | \Psi^\lambda \rangle
+ E_\text{H}^{sr,\lambda}[\rho] + E_\text{xc}^{sr,\lambda}[\rho],    \label{eq:hkf}
\end{equation}
where the first term defines the long-range universal functional $F^{lr,\lambda}[\rho]$, 
the second term $E_\text{H}^{sr,\lambda}[\rho]$ is the short-range Hartree functional, and the third term
$E_\text{xc}^{sr,\lambda}[\rho]$ is the short-range exchange-correlation functional. 
The total energy from eq~\ref{eq:es}, therefore, can be rewritten as
\begin{eqnarray}
E_\text{tot}[\rho] =& \underset{\rho \to N}{\min} \Bigl\{ F^{lr,\lambda}[\rho] + 
E_\text{H}^{sr,\lambda}[\rho] + E_\text{xc}^{sr,\lambda}[\rho] \\ \nonumber
&+\int d\mathbf{r} v(\mathbf{r}) \rho(\mathbf{r})   
\Bigr\}.
\label{eq:mdwf}
\end{eqnarray}



To formulate a self-consistent theory including a long-range exchange and correlation energies 
coming from a Green's function method, we redefine the long-range functional $F^{lr,\lambda}[\rho]$ as the 
following functional of a single-particle Green's function $G$
\begin{equation}
F^{lr,\lambda}[\rho] = \underset{G \to \rho}{\min}  \{ T[G] + E_{ee}^{lr,\lambda}[G] \}.
\end{equation}
Here, $T[G]$ is the kinetic energy functional and $E_{ee}^{lr,\lambda}[G]$ is the long-range interaction functional
of a single-particle Green's function. The search is performed over all single-particle Green's functions
yielding a given density $\rho(\mathbf{r})$. Consequently, we can write the ground state electronic energy 
as a functional of a single-particle Green's function
\begin{widetext}
\begin{eqnarray}
E_\text{tot}[\rho] & = & \underset{\rho \to N}{\min} \Bigl\{ \underset{G \to \rho}{\min} \{ T[G] + E_{ee}^{lr,\lambda}[G] \} +
E_\text{H}^{sr,\lambda}[\rho] + E_\text{xc}^{sr,\lambda}[\rho] +
\int d\mathbf{r} v(\mathbf{r}) \rho(\mathbf{r})
\Bigr\} \\ \nonumber
& = & \underset{G \to N}{\min}  \Bigl\{ T[G] + E_{ee}^{lr,\lambda}[G]  +
E_\text{H}^{sr,\lambda}[\rho] + E_\text{xc}^{sr,\lambda}[\rho] +
\int d\mathbf{r} v(\mathbf{r}) \rho(\mathbf{r})
\Bigr\},     \label{eq:min}
\end{eqnarray}
\end{widetext}
where the electron density $\rho(\mathbf{r})$ 
is calculated from the Green's function $G \to \rho$.
Note that the single-particle Green's function minimizing eq~\ref{eq:min} yields both 
the exact electron density $\rho$ and  proper total number of electrons $N$.
Therefore, we can define the total energy functional as
\begin{eqnarray}
E_\text{tot}[G] = & T[G] + E_{ee}^{lr,\lambda}[G]  +
E_\text{H}^{sr,\lambda}[G] + E_\text{xc}^{sr,\lambda}[\rho] \\ \nonumber
&+\int d\mathbf{r} v(\mathbf{r}) \rho(\mathbf{r}).   \label{eq:ef1} 
\end{eqnarray}
The long-range electron-electron interaction energy can be
decomposed into the Hartree long-range energy and the long-range exchange-correlation energy
\begin{equation}
E_{ee}^{lr,\lambda}[G] = E_\text{H}^{lr,\lambda}[G] + E_\text{xc}^{lr,\lambda}[G].
\end{equation}
The short-range and long-range Hartree energies can be folded into one term describing the all-range
Hartree energy $E_\text{H}[\rho]$. 
This leads to the following expression for the energy functional defined in eq~\ref{eq:ef1}
\begin{eqnarray}
E_\text{tot}[G] = &T[G] + E_\text{H}[\rho] + E_\text{xc}^{sr,\lambda}[G] + E_{xc}^{lr,\lambda}[G] \\ \nonumber
&+ \int d\mathbf{r} v(\mathbf{r}) \rho(\mathbf{r}).   \label{eq:ef2}
\end{eqnarray}
This energy functional (that depends on a Green's function) provides an exact decomposition of the total 
energy into short-range and long-range
components. In particular, there is no double counting of correlation effects.
The minimization of this functional with respect to a single-particle Green's function yields the ground 
state energy. It should be noted that with the exact long-range Green's function method and exact
short-range density functional the minimization of eq~\ref{eq:ef2} will produce the exact ground state
electronic energy for all possible range separation parameters $\lambda$.

In practical calculations of realistic systems, both the short-range and long-range methods
must be approximated. When employed in a range-separated framework, the standard density functional
approximations are modified to describe short-range interactions. 
The short-range exchange-correlation energy is calculated as
\begin{equation}
E_\text{xc}^{sr,\lambda} = \int d\mathbf{r}\rho(\mathbf{r}) \epsilon_\text{xc}^{sr,\lambda}(\rho),   \label{eq:exc}
\end{equation}
where $\epsilon_\text{xc}^{sr,\lambda}(\rho)$ is the short-range exchange-correlation energy
density. 
The short-range LDA exchange energy density $\epsilon_{\text{x},\sigma}^{sr,\lambda}(\rho)$ can be derived 
from the exchange hole of the homogeneous electron
gas interacting with a short-range electron-electron interaction potential~\cite{Toulouse:ijqc/100/1047}. 
Its functional form depends on the choice of the function $f(\lambda r)$~\cite{Toulouse:pra/70/062505} and 
for the error function the short-range exchange energy density 
$\epsilon_{\text{x},\sigma}^{sr,\lambda}(\rho)$ is given by~\cite{Iikura:jcp/115/3450}
\begin{widetext}
\begin{equation} \label{eq:srldax}
\epsilon_{\text{x},\sigma}^{sr,\lambda}(\rho) =  -  \frac{1}{2}\left( \frac{3}{4\pi}\right)^{1/3} 
\rho_\sigma^{1/3}(\mathbf{r}) \left( 1-\frac{8}{3}a_\sigma \left[ \sqrt{\pi}\text{erf} \left( \frac{1}{2a_\sigma}\right)  +  \left(2a_\sigma - 4a_\sigma^3\right) \exp \left( -\frac{1}{4a_\sigma^2} \right )  -   3a_\sigma+4a_\sigma^3 \right] \right),  
\end{equation}
\end{widetext}
where $a_\sigma=\lambda /(2k_{F,\sigma})$, $k_\sigma$ is the Fermi momentum given by $k_{F,\sigma}=(6\pi^2 \rho_\sigma)^{1/3}$ and $\sigma=\alpha,\beta$ is the spin index. This approximation reduces to the standard LDA 
exchange energy density at $\lambda=0$ and has a correct asymptotic
expansion for $\lambda\to \infty$~\cite{Toulouse:pra/70/062505}. Thus, it provides an interpolation between 
LDA and the correct limit as $\lambda\to \infty$. LDA was shown to be exact at 
the short-range~\cite{Gill:mp/88/1005} and, when combined with the many-body perturbation theory, such a hybrid 
method is expected to give an improved description of the dynamical correlation both in comparison to LDA 
and the perturbation theory.  Consequently, in this case LDA is used to recover a fraction of the dynamical
correlation that is missing in the finite order of perturbation theory. In subsection~\ref{sec:pes}, we provide results supporting 
this discussion  by investigating dynamical correlation in diatomic molecules. 

In order to calculate the short-range correlation energy
density $\epsilon_c^{sr,\lambda}(\rho)$, we adopted a
 scheme 
based on the following rational approximant~\cite{Toulouse:ijqc/100/1047,GoriGiorgi:pra/73/032506} 
\begin{equation}
\epsilon_c^{sr,\lambda}(r_s)=\frac{\epsilon_c(r_s)}
{1+c_1(r_s)\lambda+c_2(r_s)\lambda^2},    \label{eq:ec}
\end{equation}
where $\epsilon_c(r_s)$ is the correlation energy density for the standard Coulomb interactions ($\lambda=0$)
evaluated for the Wigner--Seitz radius $r_s(\rho)=(3/(4\pi \rho))^{1/3}$ with
$\rho(\mathbf{r}) =\rho_\alpha(\mathbf{r})+\rho_\beta(\mathbf{r})$. Equation~\ref{eq:ec} provides a way to interpolate between $\lambda=0$ and $\lambda \to \infty$ limits
and is applicable not only for the interpolation of the correlation energy density but can also be used for 
the exchange energy density~\cite{Toulouse:pra/70/062505}. Particular forms of $c_1(r_s)$ and $c_2(r_s)$ 
depend on the quantity interpolated. In this work, we used $c_1(r_s)$ and $c_2(r_s)$ 
determined by Toulouse \textit{et. al.} by analytical parameterization of the long-range correlation 
energy density from CCD and Fermi-hypernetted-chain calculations of the uniform electron 
gas~\cite{Toulouse:ijqc/100/1047}. 
The short-range correlation energy density  was then calculated as a difference 
between all-range and
long-range correlation energy densities. In this work, we have investigated two local density
approximations for the correlation energy:
Vosko--Wilk--Nusair (VWN5) functional (``form V'' parametrization in ref~\citenum{Vosko:cjp:58/1200}) as 
well as the Perdew and Wang (PW92) functional~\cite{Perdew:prb/45/13244}. PW92 uses the same spin-interpolation formula as the VWN functional but employs different expressions for
the paramagnetic correlation energy density and the ferromagnetic correction to it.
 After performing several test 
calculations, we noticed that total energies from the short-range VWN5 functional were 
within 1 kcal$\cdot$mol$^{-1}$ of those of the short-range PW92 functional. 
Consequently, we proceeded by using short-range VWN5 functional and all results reported in this work 
were obtained with it.

Having discussed theoretical background behind short-range density functionals and our specific 
choices, we
now turn to the discussion of the long-range electron-electron interaction energy. 
The long-range exchange energy 
is defined exactly in terms of the Fock exchange integral as
\begin{equation}
E_\text{x}^{lr,\lambda}=-\frac{1}{2}\sum_\sigma \int d\mathbf{r}\int d\mathbf{r}' 
\frac{|\gamma_\sigma(\mathbf{r},\mathbf{r'})|^2
\text{erf}(\lambda |\mathbf{r}-\mathbf{r'}|)}{|\mathbf{r}-\mathbf{r'}|}, \label{eq:lx}
\end{equation} 
where $\gamma_\sigma(\mathbf{r},\mathbf{r'})$ is the one-electron reduced density matrix. 
Note that the incorporation of the screening provided by the error function leads to a faster decaying 
long-range exchange contribution and, especially for metallic systems,  can result in reducing the 
computational cost~\cite{Heyd:jcp/118/8207}.

In this work, we propose to calculate the long-range correlation 
energy using single-particle Green's function methods. 
In a Green's function formalism, it is possible to correct a zeroth order Green's function $\mathbf{\mathcal{G}}(\omega)$ (which in certain cases can be a non-interacting Green's function) using the Dyson equation~\cite{fetter2003quantum} 
\begin{equation}
\mathbf{G}_\sigma(\omega) = \left[ \mathbf{\mathcal{G}}_\sigma(\omega)^{-1} - \mathbf{\Sigma}_\sigma(\omega) \right]^{-1}, \label{eq:dyson}
\end{equation}
where $\mathbf{\Sigma}_\sigma(\omega)$ is the self-energy of the system. The self-energy is an effective 
single-particle potential 
that incorporates all many-body effects present in the system. 
At this point, a connection to the density functional theory can be made. The frequency-dependent self-energy
$\mathbf{\Sigma}(\omega)$ shares some similarities with the exchange-correlation potential of DFT 
$v_\text{xc}(\rho)$ since  $v_\text{xc}(\rho)$ also connects interacting and non-interacting systems.
However, we stress that unlike $v_\text{xc}(\rho)$ in Kohn--Sham DFT, the self-energy is a dynamic, 
nonlocal and orbital-dependent quantity. This implies that a treatment of such potentials is beyond 
the Kohn--Sham scheme and it requires the so-called generalized Kohn--Sham framework
(GKS)~\cite{Seidl:prb/53/3764}. 

Calculating either the exact exchange-correlation potential or  the exact self-energy is an 
inconceivably complicated task. Fortunately, a hierarchy of 
systematically improvable approximations to the self-energy is provided by the
many-body perturbation theory~\cite{jishi2013,stefanucci2013nonequilibrium}. Examples of
such approaches include GF2,
GW~\cite{Hedin:pr/139/A796,Aryasetiawan:rpp/61/239}, and
FLEX~\cite{Bickers:prl/62/961,Bickers:ap/193/206} approximations. 

Since both the long-range exchange (eq~\ref{eq:lx}) and the long-range correlation energy 
(eq~\ref{eq:gm}) should be calculated  self-consistently with their short-range counterparts, 
it is important that such a self-consistent evaluation can be carried out easily. 
Moreover, for Green's function methods, only fully iterative schemes respect
the conservation laws and ensure that quantities obtained
by a thermodynamic or coupling constant integration from
non-interacting limits are consistent~\cite{Baym:pr/124/287,Baym:pr/127/1391}

This is why in our work, we did not employ 
any real axis single-particle Green's functions $\mathbf{G}(\omega)$ that are rational
functions in the complex plane. The rational structure of $\mathbf{G}(\omega)$ implies the 
existence of poles, for which, iterative
algorithms require pole shifting techniques~\cite{Lu:prb/90/085102,Neck:jcp/115/15,Peirs:jcp/117/4095}.
Consequently, the real axis Green's functions methods are known to present problems during 
self-consistent schemes.


We are employing an imaginary axis, single-particle Green's function  $\mathbf{G}(i\omega_n)$ that is a 
smooth function of the imaginary  argument  $i\omega_n$ and is used to describe single-particle properties 
of a statistical ensemble.  Due to the smooth structure, $\mathbf{G}(i\omega_n)$ is a convenient quantity 
for self-consistent calculations. The imaginary frequency (Matsubara) Green's function 
$\mathbf{G}(i\omega_n)$ is expressed on a grid of imaginary frequencies
located at $i\omega_n=i(2n+1)\pi/\beta$~\cite{Matsubara:ptp/14/351}, where $n=0,1,2,..$, $\beta=1/(k_\text{B}T)$ 
is the inverse temperature, $k_\text{B}$ is the Boltzmann constant and $T$ is the physical
temperature.
Providing that the imaginary frequency self-energy and Green's function were 
self-consistently determined~\cite{stefanucci2013nonequilibrium}, the 
long-range correlation energy can be calculated using the Galitskii--Migdal
formula~\cite{Galitskii:jetp/34/139}
\begin{eqnarray}
E_\text{c}^{lr,\lambda} & = & k_B T \sum_n \text{Re} \left[ \text{Tr} 
\left[ \mathbf{G}^{\lambda}_\alpha(i\omega_n)\mathbf{\Sigma}^{lr,\lambda}_\alpha(i\omega_n) \right.\right.\nonumber \\
& + & \left.\left.\mathbf{G}^{\lambda}_\beta(i\omega_n)\mathbf{\Sigma}^{lr,\lambda}_\beta(i\omega_n)
\right]\right].   \label{eq:gm}
\end{eqnarray}
We have presented equations for calculating long-range exchange (eq~\ref{eq:lx}) and long-range 
correlation energies (eq~\ref{eq:gm}), however, as we mentioned before, 
is important that they are calculated self-consistently 
with their short-range counterparts. 

Here, we outline an algorithm that allows us to perform such a self-consistent evaluation. 
It should be noted that
the formalism presented in this work is general and not limited to a specific choice of the Green's 
function method and the density functional approximation.

\begin{enumerate}
\item The calculation begins with an initial guess for the density matrix $\mathbf{P}$. For all calculations
presented in this work, the Hartree--Fock density matrix was used for this purpose. 
The method is, however, reference-independent and different
choices of the initial density matrix are possible and the same converged solution should be reached irrespective of the starting point.
\item \label{loop} The electron density is calculated using a finite set of $L$ basis functions 
$\{\phi_i(\mathbf{r})\}$ 
\begin{equation}
\rho_\sigma(\mathbf{r}) = \sum_{ij}^L P^\sigma_{ij}\phi_i(\mathbf{r})\phi_j(\mathbf{r}).
\end{equation}
\item 
The density matrix is used to calculate the all-range Hartree contribution to the Fock matrix 
according to
\begin{equation}
J_{ij} = \sum_{kl} \left(P^\alpha_{kl} + P^\beta_{kl} \right)v_{ijkl},
\end{equation}
where $v_{ijkl}$ are unscreened two-electron integrals 
\begin{equation}
v_{ijkl} = \int d\mathbf{r} \int d\mathbf{r}' \frac{\phi_i^*(\mathbf{r}) \phi_j(\mathbf{r}) \phi_k^*(\mathbf{r}')
\phi_l(\mathbf{r}')}{|\mathbf{r} - \mathbf{r}'|}.
\end{equation}
\item The short-range exchange-correlation energy is calculated using
eqs~\ref{eq:exc},~\ref{eq:srldax},~\ref{eq:ec} and the corresponding contributions to the Fock matrix are
given by
\begin{eqnarray}
\left[V_{\text{x},\sigma}^{sr,\lambda}\right]_{ij} & = & \int d\mathbf{r} v_{\text{x},\sigma}^{sr,\lambda}(\rho) \phi_i(\mathbf{r}) \phi_j(\mathbf{r}), \nonumber \\
\left[V_{\text{c},\sigma}^{sr,\lambda}\right]_{ij} & = & \int d\mathbf{r} v_\text{c}^{sr,\lambda}(\rho) \phi_i(\mathbf{r}) \phi_j(\mathbf{r}), 
\end{eqnarray}
where the short-range exchange $v_{\text{x},\sigma}^{sr,\lambda}(\rho)$ and short-range
correlation $v_\text{c}^{sr,\lambda}(\rho)$ potentials are functional
derivatives of short-range exchange and short-range correlation  functionals
with respect to the electron density:
$v_{\text{x},\sigma}^{sr,\lambda}(\rho)=\delta E_\text{x}^{sr,\lambda}[\rho]/\delta \rho_\sigma(\mathbf{r})$
and $v_{\text{c},\sigma}^{sr,\lambda}(\rho)=\delta E_\text{c}^{sr,\lambda}[\rho]/\delta \rho_\sigma(\mathbf{r})$, respectively.
\item\label{step:gf} Each of the spin components of the non-interacting Matsubara Green's function is then built according to
\begin{equation}
\mathbf{\mathcal{G}}_\sigma(i\omega_n) = \left[(i\omega_n + \mu_\sigma)\mathbf{S} - 
\mathbf{F}_\sigma \right]^{-1},
\end{equation}
where $\mu_\sigma$ is the chemical potential, $\mathbf{S}$ is the overlap matrix and $\mathbf{F}_\sigma$ is the 
Fock matrix containing all-range Hartree and short-range exchange-correlation parts
\begin{equation}
\mathbf{F}_\sigma = \mathbf{H}^\text{core} + \mathbf{J} + \mathbf{V}_{\text{x},\sigma}^{sr,\lambda} + \mathbf{V}_{\text{c},\sigma}^{sr,\lambda},
\end{equation}
where $\mathbf{H}^\text{core}$ is the core Hamiltonian matrix
\begin{equation}
H_{ij}^\text{core} = \int d\mathbf{r} \phi_i^*(\mathbf{r}) \left(-\frac{1}{2}\nabla^2_{\mathbf{r}} +
v(\mathbf{r}) \right)\phi_j(\mathbf{r})
\end{equation}
and $v(\mathbf{r})$ is the external potential.
\item The Green's function from step~\ref{step:gf} is then used to generate either the long-range 
self-energy $\mathbf{\mathcal{G}}(i\omega_n) \to \mathbf{\Sigma}^{lr,\lambda}(i\omega_n)$ or directly the 
correlated Green's function
depending on a particular Green's function method used. Both quantities are needed later and the Dyson
eq~\ref{eq:dyson} is used to obtain one from the other. 
\item The long-range exchange contribution to the Fock matrix is calculated according to
\begin{equation}
K_{ij,\sigma}^{lr,\lambda}=-\sum_{kl}P_{kl}^\sigma v_{ilkj}^{lr,\lambda}.
\end{equation}
The interacting Green's function at this point reads as
\begin{equation}
\mathbf{G}_\sigma^\lambda(i\omega_n) = \left[(i\omega_n + \mu_\sigma)\mathbf{S} - \mathbf{F}_\sigma -
\mathbf{\Sigma}^{lr,\lambda}_\sigma(i\omega_n)\right]^{-1},     \label{eq:gfc}
\end{equation}
where the Fock matrix has now both terms coming from the density functional and the Green's function method
\begin{equation}
\mathbf{F}_\sigma = \mathbf{H}^\text{core} + \mathbf{J} + \mathbf{V}_{\text{x},\sigma}^{sr,\lambda} + 
\mathbf{V}_{\text{c},\sigma}^{sr,\lambda} + \mathbf{K}^{lr,\lambda}_\sigma.  \label{eq:fc}
\end{equation}
The long-range self-energy $\mathbf{\Sigma}^{lr,\lambda}_\sigma(i\omega_n)$ describes the 
dynamical (frequency-dependent) long-range  correlation.
\item The long-range correlation energy is calculated using the correlated Green's function 
$\mathbf{G}_\sigma^\lambda(i\omega_n)$
and the long-range self-energy $\mathbf{\Sigma}^{lr,\lambda}_\sigma(i\omega_n)$ according to eq~\ref{eq:gm}.
\item The total electronic energy is calculated according to
\begin{eqnarray}
E_\text{tot} = &\frac{1}{2}\text{Tr}\left[(\mathbf{H}^\text{core}+\mathbf{f}_\alpha)\mathbf{P}_\alpha
+(\mathbf{H}^\text{core}+\mathbf{f}_\beta)\mathbf{P}_\beta\right] \\ \nonumber
&+ E_\text{xc}^{sr,\lambda} + E_\text{x}^{lr,\lambda} + E_\text{c}^{lr,\lambda} ,
\end{eqnarray}
where 
\begin{equation}
\mathbf{f}_\sigma=\mathbf{H}^\text{core}+\mathbf{J}+\mathbf{K}^{lr,\lambda}_\sigma.
\end{equation}
\item The interacting Green's function is then used to update the density matrix 
\begin{equation}
\mathbf{P}_\sigma = \frac{1}{\beta}\sum_{n}e^{i\omega_n0^+}\mathbf{G}_\sigma^\lambda(i\omega_n).
\end{equation}
\item The total electronic energy, density matrix, and Green's function are checked for convergence and,
 if necessary, a new iteration is started by sending updated density matrix to step~\ref{loop}.
\end{enumerate}

\begin{figure}
\includegraphics[width=\columnwidth]{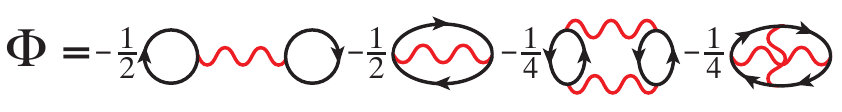}
\caption{A formal definition of the Luttinger--Ward functional as a skeleton
diagrammatic expansion, shown here for the second-order theory. Black solid lines represent
Green's functions and red wiggly lines denote electron-electron interactions
(two-electron integrals).}
\label{fig:diagr}
\end{figure}

The above algorithmic construction is in principle general and can be used in finite-temperature calculations to evaluate  the grand potential as
\begin{align}
\Omega = \Phi - \text{Tr}( \log \mathbf{G}^{-1}) - \text{Tr} (\mathbf{\Sigma} \mathbf{G}),
\end{align}
where $\Phi$ is the Luttinger--Ward (LW)~\citep{Luttinger:pr/118/1417} functional 
that is a scalar functional of a renormalized Green's function and is defined as the sum of all closed,
connected and fully dressed skeleton diagrams. The general $\Phi[{\bf G}]$ functional has the following form 
\begin{equation}
\Phi[\mathbf{G}] = E_\text{H}[\mathbf{G}] + E_\text{x}[\mathbf{G}] + E[\mathbf{G}]
\end{equation}
where $E[\mathbf{G}]$ is the correlation energy coming from frequency dependent 
$\mathbf{\Sigma}(i\omega_n)$ and $\mathbf{G}(i\omega_n)$. Since $\delta \Phi/\delta G_{ij}(i\omega_n)=\Sigma_{ij}(i\omega_n)$, 
we obtain the following expression for the self-energy
\begin{equation}\label{eq:se_all}
\mathbf{\Sigma}_\sigma = \mathbf{J} + \mathbf{K}_\sigma + \mathbf{\Sigma}_{\sigma}(i\omega_n).
\end{equation}
Application of the decomposition from eq~\ref{eq:decomp} can be understood as a splitting of interaction 
lines for every diagram leading to the following expression for the self-energy
\begin{equation}
\mathbf{\Sigma}_\sigma = \mathbf{J} + \mathbf{K}^{sr,\lambda}_\sigma + \mathbf{K}^{lr,\lambda}_\sigma+
\mathbf{\Sigma}^{sr,\lambda}_{\sigma}(i\omega_n) + \mathbf{\Sigma}^{lr,\lambda}_{\sigma}(i\omega_n).
\end{equation}
Finally, when a hybrid functional with DFT is considered, short-range exchange and short-range
correlation self-energies are approximated by static 
(frequency-independent) corresponding potentials from the density functional approximation:
$\mathbf{K}^{sr,\lambda}_\sigma \to \mathbf{V}_{\text{x},\sigma}^{sr,\lambda}$ and 
$\mathbf{\Sigma}^{sr,\lambda}_{\sigma}(i\omega_n) \to \mathbf{V}_{\text{c},\sigma}^{sr,\lambda}$ 
resulting in the following expression for the self-energy
\begin{equation}\label{eq:se_all_dft_app}
\mathbf{\Sigma}_\sigma = \mathbf{J} + \mathbf{V}_{\text{x},\sigma}^{sr,\lambda} +
\mathbf{V}_{\text{c},\sigma}^{sr,\lambda} + \mathbf{K}^{lr,\lambda}_\sigma +
\mathbf{\Sigma}^{lr,\lambda}_{\sigma}(i\omega_n),
\end{equation}
which enters the expression for the correlated Green's function shown earlier in 
eqs~\ref{eq:gfc} and~\ref{eq:fc}.
While in principle the presented formalism that merges DFT with Green's function theory is temperature dependent and completely general, in our work, we use two simplifications.
First, all practical calculations are currently limited to the $T=0$ case due to lack of reliable explicit 
finite-temperature density functional approximations.  Second, in our work, for simplicity, we have employed
the finite-temperature, self-consistent, second-order Green's function theory (GF2) for evaluating 
$\mathbf{\Sigma}_\sigma^{lr,\lambda}(i\omega_n)$. Consequently, in equations~\ref{eq:se_all} 
to \ref{eq:se_all_dft_app}, we use ${\bf \Sigma}_\sigma(i\omega_n)={\bf \Sigma}_{2,\sigma}(i\omega_n)$, ${\bf \Sigma}^{sr,\lambda}_\sigma(i\omega_n)={\bf \Sigma}^{sr,\lambda}_{2,\sigma}(i\omega_n)$, and ${\bf \Sigma}^{lr,\lambda}_\sigma(i\omega_n)={\bf \Sigma}^{lr,\lambda}_{2,\sigma}(i\omega_n)$. 
The corresponding second-order Feynman diagrams for $\Phi$ are shown in Figure~\ref{fig:diagr}.
Since for the reasons discussed above, the DFT part of calculations is carried out at $T=0$, we evaluate 
the GF2 self-energy and Green's function for large $\beta$,
corresponding to
$T\to 0$. For gapped systems, these calculations are equivalent to the $T=0$ regime.
One of the key advantages of range-separated hybrid functionals
stems from the fact that partitioning in eq~\ref{eq:decomp} is chosen such that a singularity 
is only present in the short-range operator at electron-electron coalescence, while the long-range
contribution is smooth. The absence of the singularity in the long-range part has significant 
consequences. Most importantly, a correlated method applied to the long-range electron-electron
interactions will not need to represent a cusp using a finite set of 
one-electron basis functions, thus avoiding basis sets containing functions with very high angular
momentum. In contrast to most electron correlation methods, density functionals are weakly
basis-set dependent. Therefore, range-separated hybrid functionals usually exhibit 
faster convergence of the correlation and total energies with the size of the basis set. In 
subsection~\ref{sec:basis}, we illustrate that this indeed the case for the functional presented in
this work.

In practical applications, a value of the range separation parameter $\lambda$ has to be specified before 
a calculation is carried out.
It is important that this value is chosen such that the respective approximations are evaluated within 
a regime that is optimal for their performance~\cite{Baer:arpc/61/85}.
The simplest estimation of an optimal value of $\lambda$ is based on a
local approximation~\cite{Pollet:jcp/116/1250} $\lambda(\rho)=r_s(\rho)^{-1}$. The physical motivation
behind this value is related to the fact that an electron on average occupies the sphere with boundaries 
defined by the Wigner--Seitz radius (also known as a characteristic length of the exchange). Therefore, electrons 
begin to enter an occupation sphere of the other electrons when $\lambda(\rho) \ge r_s(\rho)^{-1}$.

More sophisticated ways to find an optimal value of $\lambda$ are based on the first-principles approaches
and amount to finding $\lambda$ satisfying some relationships that an exact theory should obey. For
instance, a vertical ionization potential (IP) of a molecule containing $N$ electrons is defined as 
\begin{equation}
\text{IP}_{E(N)}^{E(N-1)} = E_\text{tot}(N-1) - E_\text{tot}(N),    \label{eq:ipdiff}
\end{equation}
where $E_\text{tot}(N)$ is the total ground state energy of a cation and $E_\text{tot}(N)$ is the total 
ground state energy of a neutral molecule. In an exact theory, $\text{IP}_{E(N)}^{E(N-1)}$ should exactly agree 
with the IP calculated from the real frequency Green's function of $N$-electron (neutral) system $\mathbf{G}_N(\omega)$.
The general idea of the  IP tuning approach is therefore to require that  IP from $\mathbf{G}_N(\omega)$ 
is as close as possible to IP calculated from
total energies of $N-1$ and $N$ electron systems. 
Therefore, an optimal value of $\lambda$ can be found by a minimization of the following bijective
function
\begin{equation}
\mathcal{T}_N(\lambda) = \bigg| \text{IP} \left[ \mathbf{G}_N^\lambda(\omega) \right] 
- \text{IP}_{E(N)}^{E(N-1)} \bigg|, \label{norm}
\end{equation}
where $\text{IP} \left[ \mathbf{G}_N^\lambda(\omega) \right]$ is the ionization potential calculated 
from the real frequency Green's function for a given value of $\lambda$.
The minimum of $\mathcal{T}_N(\lambda)$ defines an optimal $\lambda$ for which the ionization 
potential from a Green's function calculated for $N$-electron system is the closest to 
the ionization potential calculated from total energies of $N-1$ and $N$ electron systems. 
It is important to emphasize that such tuning procedure does not require any empirical input.

Several methods of calculating IP from a single-particle Matsubara Green's function of an $N$-electron 
system including the extended Koopmans theorem 
(EKT)~\cite{Smith:jcp/62/113,Day:jcp/62/115,Matos:ijqc/31/871,Morrison:jcp/96/3718} have been proposed. In this work, we adopted the following approach. 
First, the converged Fock matrix $\mathbf{F}$ coming from the imaginary axis GF2 calculation is 
transformed to the canonical representation $\mathbf{\mathcal{E}}$.  
Then the real frequency Green's function is constructed according to
\begin{equation}
\mathbf{G}(\omega) = \left[ \omega + \mu - \mathbf{\mathcal{E}} \right]^{-1},
\end{equation}
where $\omega$ is the real frequency grid point.
Then the second-order self-energy on the real frequency axis is calculated as follows~\cite{szabo1989modern}
\begin{eqnarray}
\Sigma_{ij}(\omega) =& \frac{1}{2} \sum_{ars} 
\frac{\langle rs || ia \rangle \langle ja || rs \rangle}{\omega + \mathcal{E}_a - \mathcal{E}_r - \mathcal{E}_s}\\ \nonumber
&+ \frac{1}{2} \sum_{abr} 
\frac{\langle ab || ir \rangle \langle jr || ab \rangle}{\omega + \mathcal{E}_r - \mathcal{E}_a - \mathcal{E}_b},
\end{eqnarray}
where $i,j$ denote both occupied and virtual spin orbitals, $a,b$ denote the occupied spin
orbitals only, and $r,s$ label virtual spin orbitals, $\langle rs || ia \rangle$ are the
antisymmetrized two-electron integrals. Occupied and virtual orbitals are 
defined with respect to the Hartree--Fock determinant. The self-energy is then used to
construct an updated real frequency Green's function according to 
\begin{equation}
\mathbf{G}(\omega) = \left[ \omega + \mu - \mathbf{\mathcal{E}} - 
\mathbf{\Sigma}(\omega) \right]^{-1}.
\end{equation}
The spectral function $\mathbf{A}(\omega)$ is then evaluated using
\begin{equation}
\mathbf{A}(\omega) = -\frac{1}{\pi} \text{Im} \mathbf{G}(\omega).
\end{equation}
All peaks of $\mathbf{A}(\omega)$ were shifted by the chemical potential $\mu$ and IP was
set to the closest to $\omega=0$ peak $\tilde{\omega}$ from $\omega^-$ side  
\begin{equation}\label{eq:g_ip}
\text{IP} \left[ \mathbf{G}_N^\lambda(\omega) \right] = -(\tilde{\omega} + \mu).
\end{equation}
Results of the IP-tuning approach described above are illustrated in subsection~\ref{sec:ip}.

Another constraint that an exact electronic structure theory should comply with is based on
the energy of fractional electron systems. It is well-known that the total electronic energy 
should vary linearly in the fractional electron occupancy  between 
integer electron numbers~\cite{Perdew:prl/49/1691,Perdew:pra/76/040501,Mori-Sanchez:jcp/125/201102}. Inexact methods satisfy this condition only approximately. 
To the extent that a method deviates from this condition
such a method possesses the many-electron self-interaction error. 
We have investigated this condition on the example of a two-electron system. Results are
presented and discussed in subsection~\ref{sec:sie}.

\section{Computational details}
\label{sec:comp}
In this work, we present the range-separated hybrid functional 
srSVWN5---lrGF2 that combines the SVWN5 density functional with the self-consistent second-order perturbative many-body 
Green's function method (GF2). 
In GF2, the long-range second-order self-energy is calculated in the imaginary time domain according 
to~\cite{Phillips:jcp/142/194108}
\begin{widetext}
\begin{eqnarray}
\left[\Sigma_{\alpha}^{lr,\lambda}(\tau)\right]_{ij} & = & -\sum_{klmnpq}
\left[G^\lambda_\alpha(\tau)\right]_{kl}
\left[G^\lambda_\alpha(\tau)\right]_{mn}
\left[G^\lambda_\alpha(-\tau)\right]_{pq}
v_{ikmq}^{lr,\lambda} \left( v_{ljpn}^{lr,\lambda} - v_{pjln}^{lr,\lambda} \right) \nonumber \\
&  & -\left[G^\lambda_\alpha(\tau)\right]_{mn} \left[G^\lambda_\beta(\tau)\right]_{kl}
\left[G^\lambda_\beta(-\tau)\right]_{pq}v_{ikmq}^{lr,\lambda}v_{ljpn}^{lr,\lambda}, \nonumber \\
\left[\Sigma_{\beta}^{lr,\lambda}(\tau)\right]_{ij} & = & -\sum_{klmnpq}
\left[G^\lambda_\beta(\tau)\right]_{kl}
\left[G^\lambda_\beta(\tau)\right]_{mn}
\left[G^\lambda_\beta(-\tau)\right]_{pq}
v_{ikmq}^{lr,\lambda} \left( v_{ljpn}^{lr,\lambda} - v_{pjln}^{lr,\lambda} \right) \nonumber \\
&  & -\left[G^\lambda_\beta(\tau)\right]_{mn} \left[G^\lambda_\alpha(\tau)\right]_{kl}
\left[G^\lambda_\alpha(-\tau)\right]_{pq}v_{ikmq}^{lr,\lambda}v_{ljpn}^{lr,\lambda},   \label{eq:gf2}
\end{eqnarray}
\end{widetext}
where $G_{kl}^{\lambda}(\tau)$ is the imaginary time Green's function.
The algorithm outlined above has been implemented using a locally modified version of the 
\textsc{DALTON}~\cite{dalton} program for the calculation of long-range two-electron integrals and the short-range SVWN5
exchange-correlation energy and exchange-correlation potential. An in-house GF2 code~\cite{Phillips:jcp/140/241101} 
was used to perform the self-consistent procedure and to calculate the long-range second-order self-energy.
The imaginary time Green's function and self-energy that were optimized for realistic systems were evaluated using the Legendre 
representation~\cite{Kananenka:jctc/12/564}
and a cubic spline interpolation algorithm~\cite{Kananenka:jctc/12/2250} was employed to optimize 
imaginary-frequency quantities. The convergence of the total energy with respect to the size of the Legendre
expansion, imaginary time and imaginary frequency grids was verified. Total electronic energies were converged 
to 5$\cdot$10$^{-6}$ au The inverse temperature was set to $\beta=100$ au, corresponding to a physical
temperature below the excitation energy necessary to occupy the lowest unoccupied level
of all systems considered in this work. Results of 
standard methods: SVWN5, CCSD(T) and FCI, reported in this work, 
were obtained with \textsc{gaussian 09}~\cite{g09} program.

\section{Results and Discussion}
\label{sec:results}
In this section, we present and analyze numerical results of the application
of the srSVWN5---lrGF2 functional to concepts discussed above.

\subsection{Basis set convergence}
\label{sec:basis}
\begin{figure*}
\includegraphics[width=15.0cm,height=6cm]{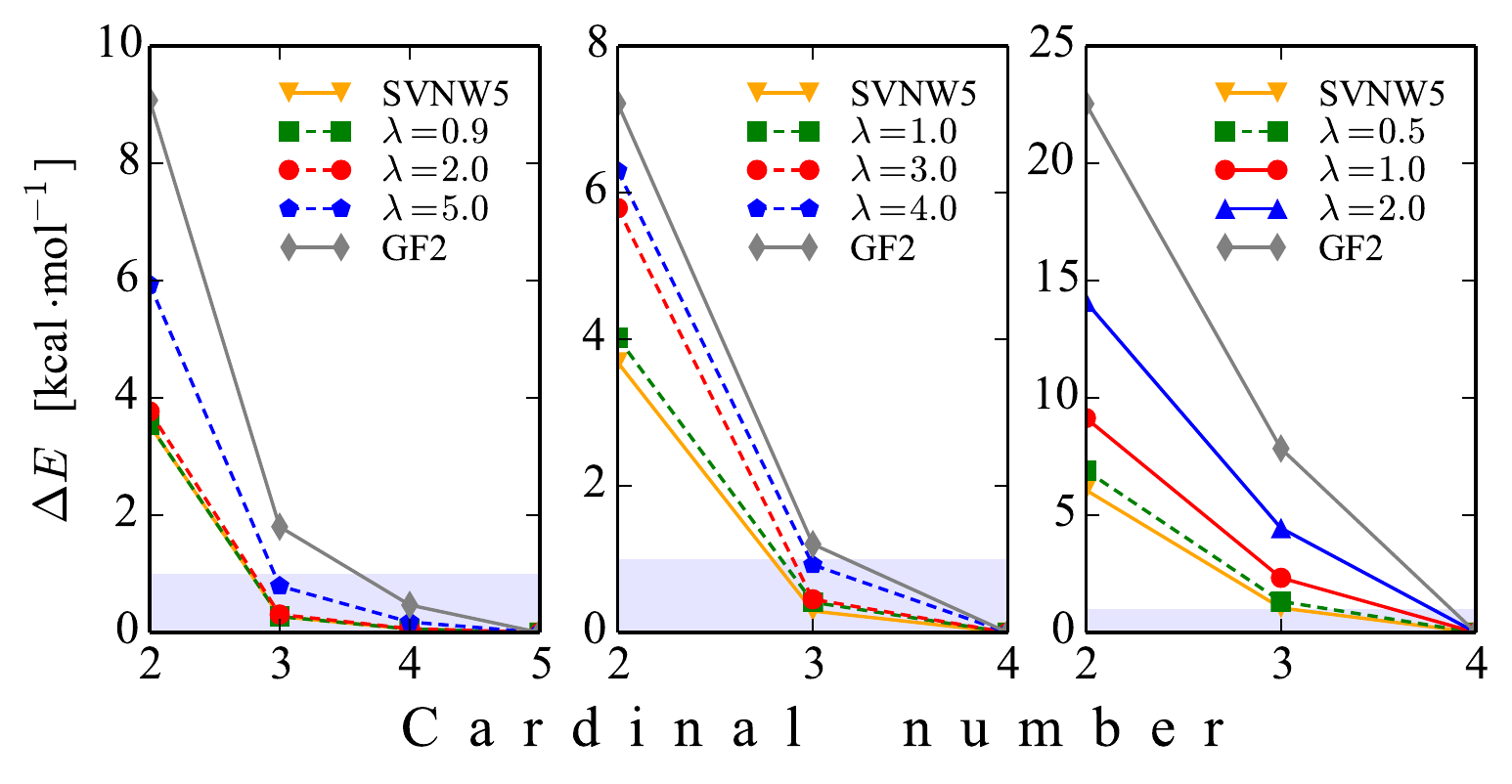}
\caption{Basis set convergence as a function of the range separation parameter $\lambda$. 
$\Delta E = |E_\text{aug-cc-pVYZ} - E_\text{aug-cc-pVXZ}|$ is plotted on the y-axis, while the cardinal 
number X is plotted on the x-axis. $\Delta E$ is given in kcal$\cdot$mol$^{-1}$. The shaded 
area shown in every plot corresponds to  1 kcal$\cdot$mol$^{-1}$.
{\em Left panel}: Results for the He atom with $X \in \{D,T,Q,5\}$, $Y=5$. {\em Middle panel}: 
Results for the H$_2$ molecule at the equilibrium bond length R(H--H)=1.4 au, 
with $X \in \{D,T,Q\}$, $Y=4$. {\em Right panel}: Results for the Mg atom with $X \in \{D,T,Q\}$, $Y=4$.}
\label{fig:basis}
\end{figure*}

In this section, for a series of aug-cc-pVXZ augmented
correlation-consistent polarization Dunning basis sets~\cite{Dunning:jcp/90/1007,Woon:jcp/100/2975,Prascher:tca/128/69}, we investigated
the convergence of the srSVWN5---lrGF2 total energy as a function
of the range separation parameter $\lambda$ for three systems: He and Mg atoms as well as
H$_2$ molecule at the equilibrium distance R(H--H) = 1.4 au. 
We studied the convergence of 
the total energy with respect to the cardinal number $X$, corresponding to the highest angular 
momentum in a given basis set $\mathcal{L}$ (note, that for He, $X=\mathcal{L}-1$). The following
values of $X$ were used: $X \in \{D,T,Q,5\}$ for He
and $X \in \{D,T,Q\}$ for H$_2$ and Mg. Relative to the total energy, obtained for a basis set with $X$=5 for He and $X$=4
for H$_2$ and Mg, the total electronic energies of the srSVWN5---lrGF2 functional are 
plotted in Figure~\ref{fig:basis}. 
In Figure~\ref{fig:basis}, SVWN5 energies, corresponding to  orange lines with triangles, confirm 
that density functional approximations converge very rapidly with respect to the
basis set size. GF2 energies, illustrated by gray lines with diamonds, result in the slowest convergence
for every system studied in this work. Any mixture of SVWN5 and GF2 leads to an improved convergence when compared to  GF2.
For $\lambda < 1$, srSVWN5---lrGF2 converges as fast as SVWN5 for all the systems considered here.
For values of $\lambda > 1$, for both H$_2$ molecule and Mg atom, the convergence of the srSVWN5---lrGF2 
functional is much slower than that one of the parent SVWN5 functional.
Filled  area shown in every plot corresponds to a difference of 1 kcal$\cdot$mol$^{-1}$ from the largest basis set used for the system.
For all three systems, SVWN5 calculations converged
within 1 kcal$\cdot$mol$^{-1}$ away from the largest basis set  for cc-pVTZ ($X$=3) basis set. For the
same basis set, the GF2 energy became almost converged only for H$_2$ molecule. 
In Figure~\ref{fig:basis}, for each of the cases analyzed, we also show the largest $\lambda$ for which 
the total energy for the cc-pVTZ basis set is 1 kcal$\cdot$mol$^{-1}$ away from the energy in the largest basis set used in that system.
It corresponds to $\lambda=5$, $\lambda=4$ and $\lambda=0.5$
for He, H$_2$ and Mg respectively. 

Similarly to wave-function methods, pure Green's function methods
converge fairly slowly with respect to the basis set size. By using the density functional method 
to describe short-range interactions a faster converge with respect to the basis set size is achieved.

\subsection{Potential energy surface of diatomic molecules}
\label{sec:pes}
The accuracy of popular density functionals around equilibrium geometries stems from a satisfactory description
of the short-range dynamical correlation.  
In this section, we illustrate the dynamical correlation in the srSVWN5---lrGF2 functional 
by analyzing dissociation curves of diatomic molecules.

\begin{figure}
\includegraphics[width=6.0cm,height=6cm]{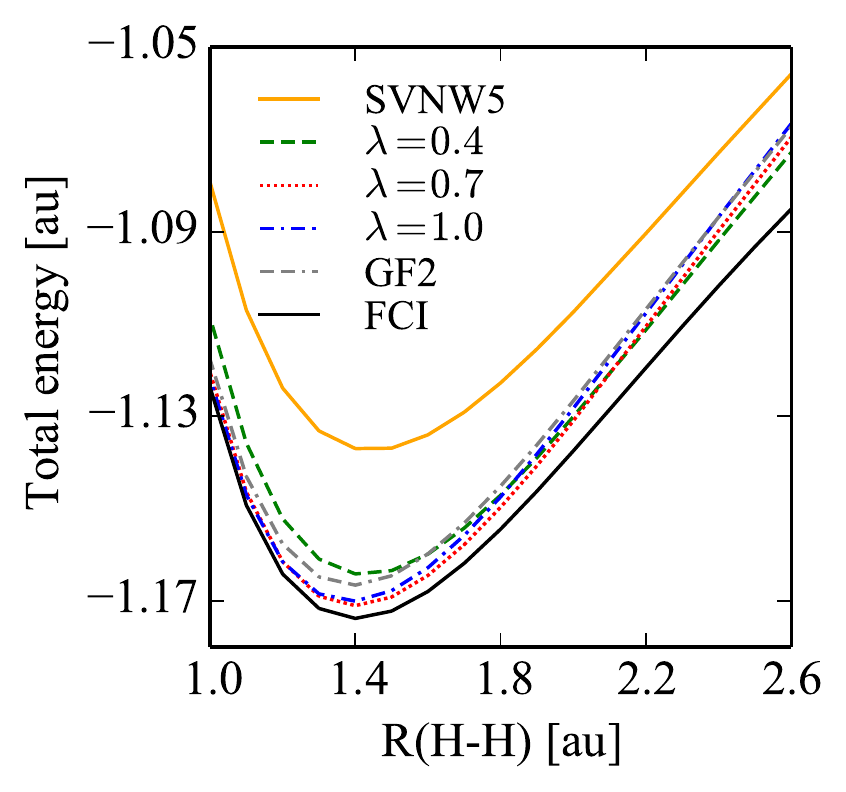}  
\caption{A dissociation curve of the H$_2$ molecule 
calculated using the srSVWN5---lrGF2 functional for different values of $\lambda$.
The SVWN5, GF2, and FCI results are shown for comparison. All calculations employed the cc-pVQZ basis set.}
\label{fig:H2_diss}
\end{figure}

First, for the H$_2$ molecule, we looked at absolute values of the total electronic energy near the 
equilibrium geometry. We performed spin-restricted 
total energy calculations using the srSVWN5---lrGF2 functional for different values of the range separation
parameter $\lambda$ scanning over values of interatomic  distances around the equilibrium geometry using the 
cc-pVQZ~\cite{Dunning:jcp/90/1007} basis set. 
The dissociation curve is illustrated in Figure~\ref{fig:H2_diss}. Full Configuration Interaction (FCI)
energies are also included and shown for comparison. 
It is clear that GF2 produces energies that are much closer to FCI than SVWN5. This suggests that GF2 recovers
the dynamical correlation better than SVWN5. However, obviously due to a finite order truncation, GF2 does not
recover all of the dynamical correlation. GF2, SVWN5, and srSVWN5---lrGF2 tend to be inaccurate far away 
from equilibrium. This is not surprising since all these methods are not well-suited for systems with a
significant strong correlation. 
As the contribution from GF2 increases (orange line $\to$ green line $\to$ cyan line, etc), the total 
energy gradually
approaches the FCI energy and when $\lambda \in \left[0.7,0.8\right]$ the total energy becomes almost 
stationary with respect 
to changes in $\lambda$. For example, E($\lambda$=0.8) $-$ E($\lambda$=0.7) = 0.1 kcal$\cdot$mol$^{-1}$. 
In particular, $\lambda=0.7$ corresponds to the best match of the
dynamical correlation coming from two respective 
approaches and produces an equilibrium distance energy which is only $1.7$ kcal$\cdot$mol$^{-1}$ 
away from FCI. For the same internuclear distance, SVWN5 and GF2 errors 
are 23.1 kcal$\cdot$mol$^{-1}$
and 4.5 kcal$\cdot$mol$^{-1}$, respectively. Overall we conclude that the short-range SVWN5 functional 
is efficient in adding the missing dynamical correlation to GF2.

\begin{figure}
\includegraphics[width=6.0cm,height=6cm]{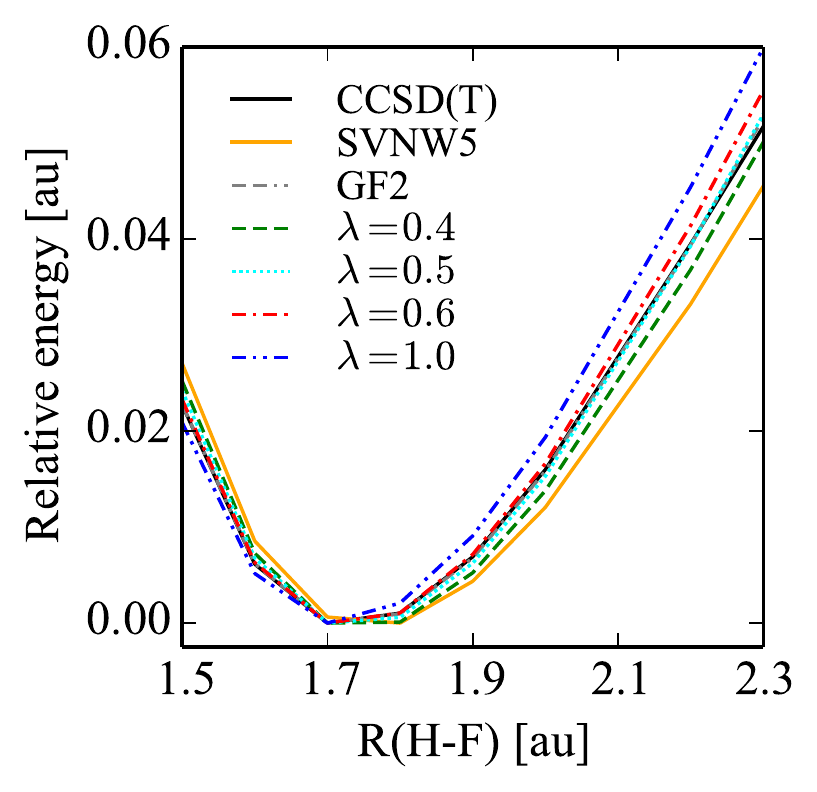}  
\caption{A dissociation curve of the HF molecule  
calculated using the srSVWN5---lrGF2 functional for different values of $\lambda$.
The SVWN5, GF2 and CCSD(T) results are shown for comparison. All calculations employed the cc-pVQZ basis set.}
\label{fig:HF_diss}
\end{figure}

The second case we considered was the dissociation of the HF molecule. Rather than looking at the absolute 
values of the electronic energy, here we focus on the electronic energies
relative to the minimum on the dissociation curve. These energies are responsible for the shape of the dissociation curve. The 
reference energies are provided by CCSD(T)~\cite{Bartlett:rmp/79/291,Raghavachari:cpl/157/479} method. 
The cc-pVQZ basis set was used
in all calculations. The results are illustrated in Figure~\ref{fig:HF_diss}.
It should be noted that the
shape of the GF2 dissociation curve is in a very good agreement with that of CCSD(T), while the SVWN5 energy grows 
too slow with the increasing internuclear separation beyond the equilibrium distance. 
Mixing GF2 and SVWN5 for small $\lambda$ up to $\lambda=0.3-0.4$ fixes this
behavior and produces the shape approaching the CCSD(T) quality. 
As $\lambda$ increases past $\lambda=0.4$, the energy 
as a function of the internuclear separation starts to grow too fast. Mixing in a larger fraction of GF2 
turns this behavior around and for $\lambda>1$, srSVWN5---lrGF2 energies start to
slowly approach GF2 energies. For internuclear 
distances up to R(H--F)=2.3 au, $\lambda=0.5$ produces relative energies closely matching those
of GF2 and CCSD(T) methods. We conclude that the srSVWN5---lrGF2 functional is able to reproduce correctly
the shape of the dissociation curve near the equilibrium geometry. The srSVWN5---lrGF2 functional does not
improve upon GF2, since GF2 being
an \textit{ab-initio}, perturbative method already correctly describes the dynamical correlations in the HF molecule. 
Nonetheless, an apparent improvement comes
from the fact that with the srSVWN5---lrGF2 functional these energies can be reached using basis sets with a
lower angular momentum when compared to standard GF2, as illustrated in the previous subsection~\ref{sec:basis}.

\subsection{IP tuning of range-separation parameter $\lambda$}
\label{sec:ip}
Following the prescription given in Section~\ref{sec:theory}, we have employed an IP-based tuning approach 
to find optimal values 
of the range separation parameter $\lambda$ for seven closed-shell 
atoms: He, Be, Ne, Mg, Ca,
 Ar, and Kr, as well as fifteen closed-shell molecules: H$_2$CO, CH$_4$, NH$_3$, N$_2$,
Li$_2$,  CO$_2$, CO, LiH, CH$_3$OH, H$_2$O$_2$, N$_2$H$_4$, H$_2$S, PH$_3$,
Na$_2$, and HCN. Experimental geometries were taken from ref~\citenum{nist}. The
cc-pVTZ~\cite{Prascher:tca/128/69,Dunning:jcp/90/1007,Wilson:jcp/110/7667,Woon:jcp/98/1358,Koput:jpca/106/9595}
basis set was used for calculations of both atoms and molecules present in this test set. 
For the cc-pVTZ and larger basis sets, the value of $\lambda$ remained constant indicating that it is converged with respect to the basis set size.

\begin{figure*}
\centering
\includegraphics[width=15.0cm,height=6cm]{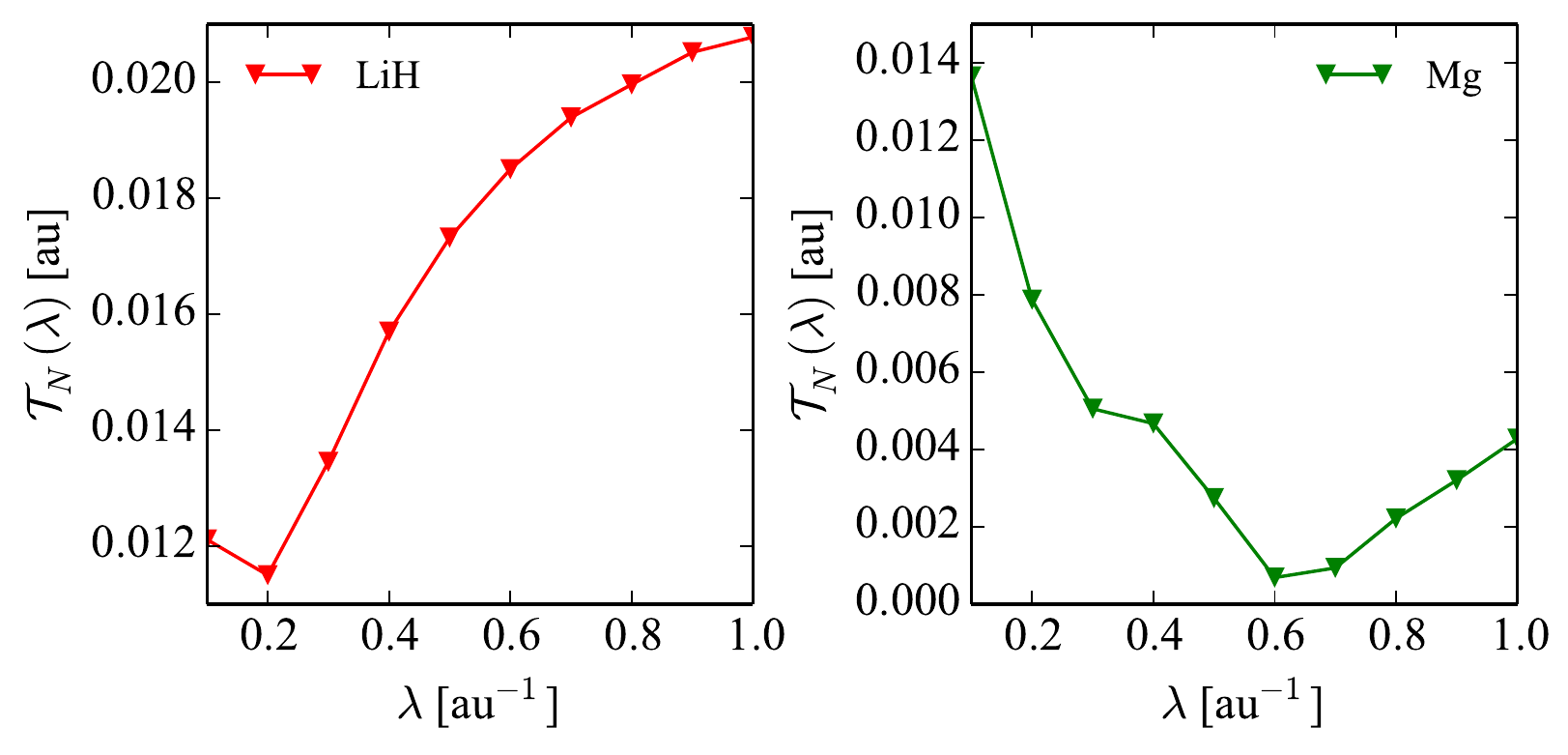} 
\caption{The absolute difference, as a function of the range separation parameter $\lambda$, between IPs 
calculated from the Green's function $\mathbf{G}_N^\lambda(\omega)$ and
IP$^{E(N-1)}_{E(N)}$ (from eq~\ref{eq:ipdiff}) using the  srSVWN5---lrGF2 
functional for LiH molecule (\textit{left panel}) and Mg atom (\textit{right panel}). 
All calculations were performed in the cc-pVTZ basis set.}
\label{fig:ip_norm}
\end{figure*}

To find an optimal value of the range separation parameter for each system in the test set, a
series of calculations were performed for $\lambda \in [0.1,1.5]$ with the step-size set to $\Delta \lambda=0.1$. 
In most cases, the $\mathcal{T}_N$ norm as a function of $\lambda$ was found to have one pronounced minimum 
that was taken as an optimal $\lambda$.
For illustration purposes, we show $\mathcal{T}_N(\lambda)$ norm for LiH molecule andMg atom in 
Figure~\ref{fig:ip_norm}. In the case of Mg atom, a very small discrepancy between two ways of calculating IP was
found for $\lambda=0.6$ with the error $\mathcal{T}_N \approx 7\cdot 10^{-4}$ while for LiH
molecule the
smallest difference between $\text{IP} \left[ \mathbf{G}_N^\lambda(\omega) \right]$ and $\text{IP}_{E(N)}^{E(N-1)}$
turned out to be larger and equal to $\mathcal{T}_N \approx 0.011$ corresponding to the optimal value of
$\lambda=0.2$.

Note that if smaller differences are desired, then a further fine-tuning of $\lambda$ can be
performed by using a root-finding algorithm such as bisection~\cite{burden1985numerical}. 
In this work, we adopted a commonly used approach and
narrowed the optimal value of $\lambda$ down to only one decimal point.
In a similar way, optimal values of the range separation parameter $\lambda$ were obtained for all systems
in this test set. 

To examine how accurately IPs can be calculated based on such an IP-tuning approach, we used the
optimally tuned srSVWN5---lrGF2 functional to calculate IPs and compared them with IPs calculated 
using standard SVWN5 and GF2 methods, as well as experiment. The experimental vertical IPs
were taken from ref~\citenum{nist}.
For consistency IPs for GF2, srSVWN5---lrGF2 with the optimal $\lambda$, and SVWN5 were calculated according to eq~\ref{eq:ipdiff}
 and listed in Table~\ref{tab:1}.
\begin{table*}
\caption{Ionization potentials (IP) calculated as $\text{IP} = E_\text{tot}(N-1) 
- E_\text{tot}(N)$ using SVWN5, GF2, and srSVWN5---lrGF2 methods. 
The cc-pVTZ basis set was employed in all calculations.
For srSVWN5---lrGF2 calculations, the optimal value of $\lambda$  is listed in the second column\textsuperscript{\emph{a}}.
}
\begin{tabular}{@{\extracolsep{4pt}}lcccccccc@{}}
\toprule
 & Opt. & \multicolumn{2}{c}{srSVWN5---lrGF2} & \multicolumn{2}{c}{GF2} & \multicolumn{2}{c}{SVWN5} & Expt. \\
   \cline{3-4} \cline{5-6} \cline{7-8}
   & $\lambda$  & IP & Error &   IP & Error &  IP & Error & \\
\multicolumn{9}{c}{Atoms}  \\
  He &  0.9 & 24.59 & 0.00  & 24.36 & 0.23 & 24.30 & 0.29 & 24.59 \\
  Be &  0.1 &  9.17 & 0.15  &  8.83 & 0.49 &  9.02 & 0.30 & 9.32 \\
  Ne &  0.5 & 22.16 & 0.60  & 21.50 & 0.06 & 22.09 & 0.53 & 21.56 \\
  Mg &  0.6 &  7.52 & 0.13  &  7.31 & 0.34 &  7.72 & 0.07 & 7.65 \\
  Ar &  0.5 & 15.97 & 0.21  & 15.66 & 0.10 & 16.08 & 0.32 & 15.76 \\
  Ca &  0.7 &  5.98 & 0.13  &  5.93 & 0.18 &  6.21 & 0.10 &  6.11  \\
  Kr &  0.5 & 14.33 & 0.33  & 14.03 & 0.03 & 14.44 & 0.44 & 14.00 \\
 \multicolumn{9}{c}{Molecules}  \\
 H$_2$CO &  0.1 & 10.98 & 0.09 & 10.86 & 0.03 & 10.88 & 0.01 & 10.89 \\
 CH$_4$  &  0.1 & 14.29 & 0.06 & 14.32 & 0.03 & 14.02 & 0.33 & 14.35 \\
 NH$_3$  &  0.8 & 10.72 & 0.10 & 10.70 & 0.12 & 11.01 & 0.19 & 10.82 \\
 N$_2$   &  0.1 & 15.66 & 0.08 & 15.15 & 0.43 & 15.58 & 0.00 & 15.58 \\
 Li$_2$  &  0.3 &  5.34 & 0.61 &  4.94 & 0.21 &  5.31 & 0.58 &  4.73 \\
 CO$_2$  &  0.1 & 14.32 & 0.55 & 13.88 & 0.11 & 13.99 & 0.22 & 13.77 \\
 CO   &  0.1 & 14.14 & 0.13 & 13.72 & 0.29 & 14.07 & 0.06 & 14.01 \\
 CH$_3$OH &  0.1 & 10.89 & 0.07 & 10.96 & 0.00 & 10.76 & 0.20 & 10.96 \\
 LiH  &  0.2 &  8.30 & 0.40 &  7.75 & 0.15 &  8.21 & 0.31 &  7.90 \\
 H$_2$O$_2$ &  0.1 & 11.46 & 0.24 & 11.19 & 0.51 & 11.40 & 0.30 & 11.70 \\
 N$_2$H$_4$ &  0.1 &  9.53 & 0.55 &  9.56 & 0.58 &  9.41 & 0.43 & 8.98  \\
 H$_2$S  &  0.5 & 10.54 & 0.04 & 10.33 & 0.17 & 10.63 & 0.13 & 10.50 \\
 PH$_3$  &  0.6 & 10.57 & 0.02 & 10.47 & 0.12 & 10.65 & 0.06 & 10.59 \\
 HCN &  1.2 & 12.09 & 0.70 & 12.90 & 0.70 & 14.04 & 0.44 & 13.60 \\
 Na$_2$ &  0.6 &  4.94 & 0.05 &  4.68 & 0.21 &  5.25 & 0.36 &  4.89 \\
 m.a.v.    &      &       & 0.24 &       & 0.23 &       & 0.26 &       \\
\end{tabular}
\textsuperscript{\emph{a} Experimental geometries and vertical IPs were taken from
NIST Computational Chemistry Comparison and Benchmark Database~\cite{nist}.}\\
\label{tab:1}
\end{table*}
It is worth noting that noble gases starting from Ne atom require the same value of $\lambda=0.5$
and, in general, moving down the periodic table leads to larger optimal values of $\lambda$.
The mean absolute errors of the srSVWN5---lrGF2 functional, the standard SVWN5 functional, and GF2 are 
0.24 eV, 0.26 eV and 0.23 eV, respectively. 
For the srSVWN5---lrGF2 functional, evaluating IP either from a Green's function (eq~\ref{eq:g_ip}) or from 
the difference between energies of $N$ and $N-1$ electron systems (eq~\ref{eq:ipdiff}) leads to the 
same results and these results are converged with respect to the basis set size. 
In contrast, for GF2, evaluating IP from eq~\ref{eq:g_ip} or eq~\ref{eq:ipdiff} leads to significantly
different results. The GF2 IPs calculated from eq~\ref{eq:g_ip} have large errors since the cc-pVTZ basis set 
is not large enough. The IPs calculated from eq~\ref{eq:ipdiff} benefit from the cancellation of the basis 
set error. Consequently, the GF2 magnitude of the error that is presented in Table~\ref{tab:1} benefits 
from fortuitous cancellations of errors. 
The benefit of using the range separated functional is in the agreement of IP when using both definitions 
and in avoiding the need of large basis sets.
As we observe from Table~\ref{tab:1}, GF2 tends to predict 
better IPs for atoms while
the srSVNW5---lrGF2 functional is more accurate for molecules. 


\subsection{Many-electron self-interaction error}
\label{sec:sie}
\begin{figure}
\centering
\includegraphics[width=6.0cm,height=6cm]{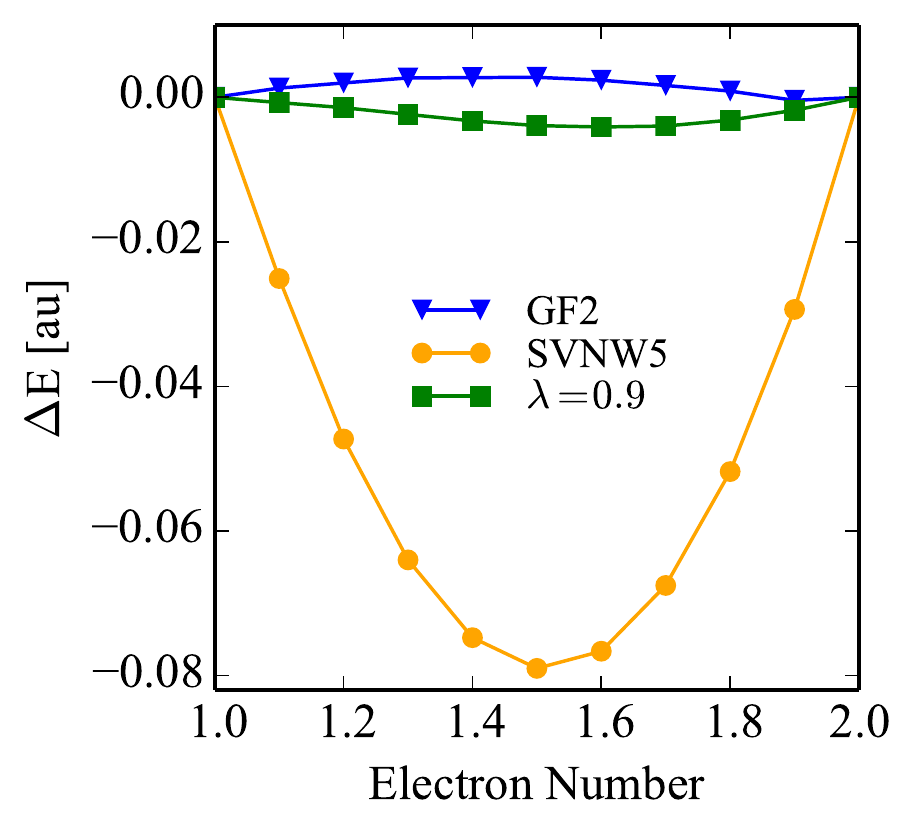} 
\caption{The energy difference $\Delta E = E^\mathcal{M} - E^\mathcal{M}_\text{lin}$ for He atom, calculated
using the srSVWN5---lrGF2 functional with the IP-tuned optimal value of $\lambda=0.9$ in comparison to that of SVWN5 and GF2
with the aug-cc-pVTZ basis set.  
$E^\mathcal{M}$ is the energy evaluated with a fractional electron number, and $E^\mathcal{M}_\text{lin}$ is the 
linear interpolation between integer electron points for  method $\mathcal{M}$.}
\label{fig:sie}
\end{figure}

The one- and many-electron self-interaction error in approximate density 
functionals originates from an incomplete 
cancellation of the spurious electrons self-repulsion by the exchange energy.
GF2 includes all the proper exchange and Hartree self-energy diagrams up to the second order
and is, therefore, one-electron self-interaction free. We have previously 
illustrated that GF2 also has a very small two-electron self-interaction error~\cite{Phillips:jcp/142/194108}. 
On the other hand, LDA, is known to have pronounced one- and many-electron self-interaction errors due to a wrong 
asymptotic decay of the exchange-correlation potential~\cite{Leeuwen:pra/49/2421}. It seems very likely that 
an application of GF2 for long-range
interactions while keeping LDA within the short range would provide an improvement over LDA by itself.
In this section, we analyze in detail the self-interaction error of the srSVWN5---lrGF2 functional. 
As we mentioned earlier, the fractional charge error is directly related to the 
self-interaction error. 
To observe it, we calculated the total electronic energy of He atom as a function of  the fractional
electron number: $N=1+\delta$ for $\delta \in [0,1]$. 
In Figure~\ref{fig:sie}, we plot the deviation from the 
linearity: $\Delta E=E^\mathcal{M}(N) - E^\mathcal{M}_\text{lin}$, where 
$E^\mathcal{M}(N)$ is the energy from
method $\mathcal{M}$ calculated for a system with $N$ electrons and $E^\mathcal{M}_\text{lin}$ is the 
linear interpolation between integer electron points for the same method $\mathcal{M}$. The IP-tuned optimal
value of $\lambda=0.9$ was used in srSVWN5---lrGF2 calculations. The aug-cc-pVTZ~\cite{Woo:jcp/100/2975} 
basis set was employed in all calculations.
It is clear from Figure~\ref{fig:sie} that GF2 has a very small fractional charge error showing a 
small concave 
behavior, therefore indicating a small localization error. SVWN5 exhibits a massive 
fractional charge error and pronounced convex 
character. This opposite behavior of SVWN5 indicates
a delocalization error common for local, semilocal, and hybrid density functionals~\cite{Cohen:cr/112/289}. 
On the other hand, srSVWN5---lrGF2 calculations for the IP-tuned range separation parameter 
$\lambda$  display only a slightly convex 
behavior and errors that are very similar 
to GF2, thus greatly improving over SVWN5. We conclude that adding a fraction of the
many-body Green's function method can significantly mitigate the self-interaction error
present in the standard density functionals. In this regard, the  srSVWN5---lrGF2 functional
is similar to popular range-separated hybrid functionals employing the exact exchange for long-range
interactions.

\subsection{Locality of self-energy}
\label{sec:locality}
\begin{figure}
\centering
\includegraphics[width=7.0cm,height=12cm]{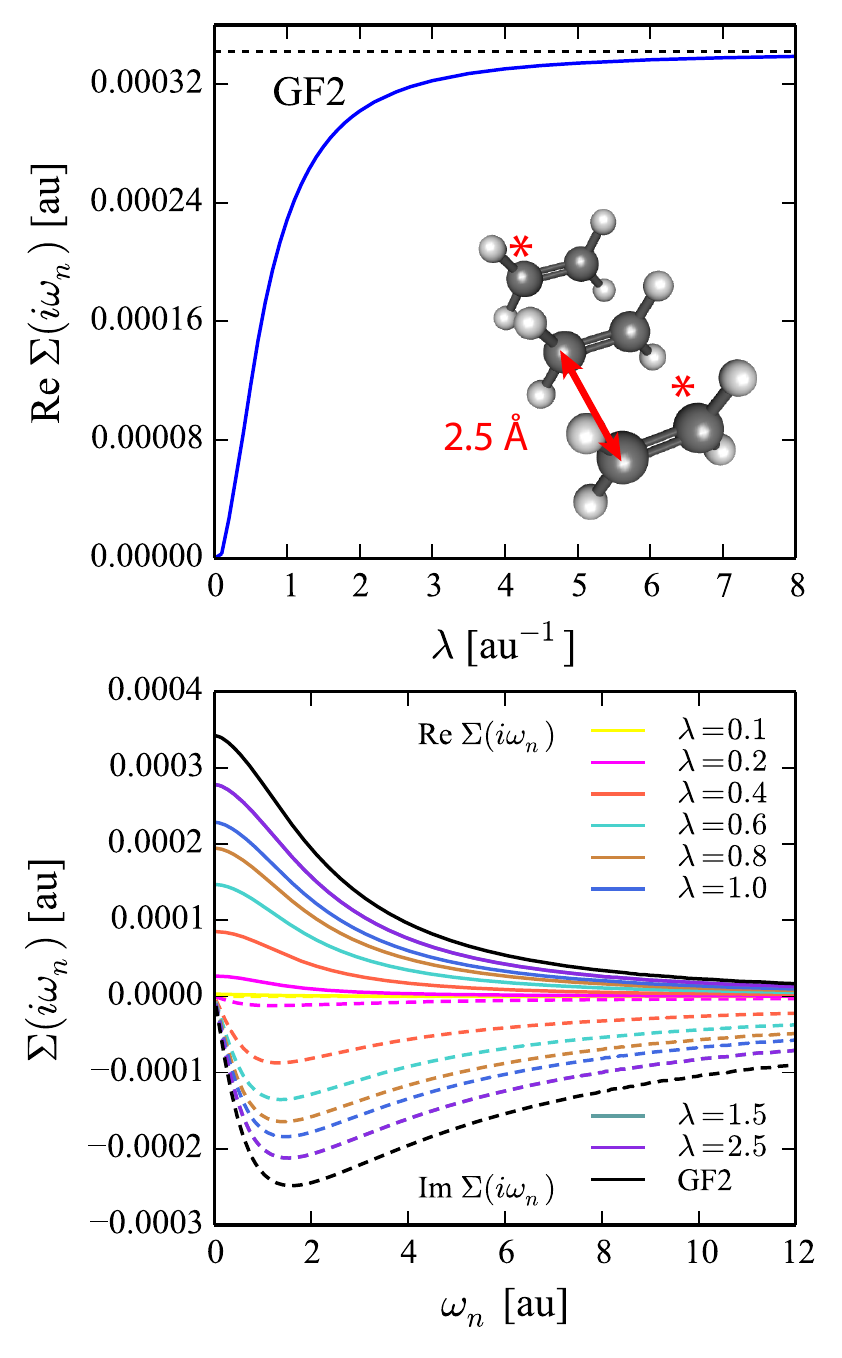} 
\caption{\textit{Top panel}: The real part of the srSVWN5---lrGF2 self-energy matrix element between two carbon atoms denoted by red stars
for the $n=0$ imaginary frequency as a function of the range separation parameter 
$\lambda$ for three ethylene molecules arranged as shown in the inset.
\textit{Bottom panel}: Both real (solid lines) and imaginary (dashed lines) parts
of the self-energy as a function of the imaginary frequency calculated for different values of $\lambda$ (bottom). All calculations are with DZP basis set.}
\label{fig:loc}
\end{figure}
In this section, we discuss implications of using range-separated hybrid functionals for the self-energy.
It is expected that by varying $\lambda$ the magnitude of self-energy can be gradually changed.
To illustrate this, srSVWN5---lrGF2 calculations were performed for three ethylene molecules, each at the 
experimental geometry~\cite{nist}, placed 2.5 \AA apart from each other (see the top panel of Figure~\ref{fig:loc}).
A matrix element of the imaginary frequency self-energy
between 2\textit{p} orbitals of the two most distant carbon atoms, denoted by red stars on the top panel of
Figure~\ref{fig:loc}, is calculated as a function of $\lambda$ using the DZP~\cite{Dunning:jcp/53/2823} basis set.
Real and imaginary parts of  
self-energy for different $\lambda$ values are shown in the bottom panel of Figure~\ref{fig:loc} using
solid and dashed lines, respectively. Colors from the lightest to the darkest correspond to an increasing fraction of GF2.
The self-energy increases most rapidly for small values of  $\lambda$, up to $\lambda \approx 0.7-0.9$, then it begins 
to slowly converge to the GF2 self-energy.
To see it more clearly, in the top panel of Figure~\ref{fig:loc}, we plotted the real part of self-energy for 
$n=0$ Matsubara frequency. It grows most rapidly for the small fractions of GF2. Overall this behavior resembles
the error function which is used to scale the 
two-electron integrals to obtain the
long-range terms. The possibility to arbitrarily 
scale the self-energy in the range separated approach has 
important consequences. For example, the srSVWN5---lrGF2 calculation is less computationally demanding comparing
to the standard GF2 calculation since 
the evaluation of the self-energy according to
eq~\ref{eq:gf2} can be carried over a truncated set of orbitals due to the faster decay of its matrix elements. Additionally, using a range-separated Green's
function functional as a low-level method e.g. in self-energy embedding 
theory~\cite{Kananenka:prb/91/121111,Lan:jcp/143/241102,Lan:jctc/12/4856, Lan:jpcl/8/2200} calculations 
of periodic systems can  be beneficial since, as we demonstrated before, such hybrids require smaller basis sets than the original \textit{ab-initio} Green's function methods. Consequently, they possibly eliminate many problems such as linear dependences that happen when large, diffuse basis sets are used 
in calculations of periodic systems.
Moreover, using these hybrid approaches, the number of unit cells required to evaluate the self-energy matrix is lowered due to a faster decay of its intercell matrix elements.

\section{Conclusions and Outlook}
\label{sec:conc}
In this paper, we have discussed the theoretical framework for building a range-separated hybrid 
functional combining both DFT and Green's function methods. In principle, this framework is general and can 
be used to combine various DFT functionals and Green's function methods. In particular, to maintain the
generality of our discussion, we have focused on describing the relationship of this range-separated functional
to the Luttinger-Ward functional which is temperature dependent. Since at present, only the zero temperature
DFT functionals are well established, we executed all the practical applications of the short-range DFT --
long-range Green's function functional using the zero temperature  SVWN5 functional for the description of the
short range and the temperature dependent GF2 method setting
$T\to 0$ for the description of the long range.

We believe that the presented range-separated hybrid functional called srSVWN5---lrGF2 is interesting for two
communities. In condensed matter, among the LDA+DMFT practitioners, there has been a long-standing problem of
removing the double counting of electron correlation present when LDA is combined with the DMFT treatment
employing the Green's function methods. We believe that our presentation of the short-range DFT -- long-range
Green's function functional is directly relevant to this community and gives a rigorous prescription how to
avoid the double counting problem by employing the range separation of Coulomb integrals. Provided that the
range separation parameter $\lambda$ can be optimized based on one of the exact properties of either the DFT 
or the Green's function methods, such a range-separated hybrid provides an \textit{ab-initio} treatment of
realistic systems. 

On the other hand, the short-range DFT -- long-range Green's function hybrid functional 
is obviously relevant to the DFT community since it can be viewed as a higher rung of the 
``Jacob's ladder'' of the functionals.
Similarly to other high rungs, srSVWN5---lrGF2 employs unoccupied orbitals, is non-local, and has an explicit
frequency dependence. Provided that explicitly temperature dependent short-range DFT functionals become
established enough, the presented functional can also be made temperature dependent in a straightforward
manner. 

We have demonstrated that the functional presented in this work 
offers several attractive advantages when compared to the methods used in its construction. 
Similarly to range-separated hybrid functionals with
other many-body methods such as CI, MP2, CASCF, NEVPT2, CCSD, and RPA, srSVWN5---lrGF2 exhibits
a rapid convergence with respect to the one-electron basis set. This fast convergence with respect to the basis
set size, for the Green's function methods 
provides an additional advantage, since smaller basis sets require fewer imaginary time and imaginary
frequency grid points, resulting in reduced computational cost. Additionally, we have illustrated that
the srSVWN5---lrGF2 functional has a smaller self-interaction error when compared to the standard SVNW5 functional.
This is beneficial in calculations involving molecular thermochemistry, 
reaction barriers,
binding energy in charge transfer complexes, polarizabilities, and molecular conductance. 
Even though the standard density functionals provide an accurate description of the short-range
dynamical correlation, we have shown on the example of the HF and H$_2$ molecules that the srSVWN5---lrGF2 functional can describe the dynamical correlation even more accurately.

Moreover, we presented a 
first principles approach to finding an optimal value of the range separation parameter based on the 
calculation of ionization potentials of atoms and molecules. While the overall accuracy of the IPs evaluated using srSVWN5---lrGF2
is similar to that of GF2 evaluated as the difference between total electronic energies of 
$N$ and $N-1$ electron systems, srSVWN5---lrGF2 results are converged with respect to the basis set size
and do not rely on any fortuitous cancellation of errors. Moreover, for srSVWN5---lrGF2 evaluating the IP 
directly from the Green's function poles or using the energy difference between $N$ and $N-1$ electron 
systems results in the same answer. This is not the case for GF2 when the calculations are carried out in a
basis set that is not large enough.

We have demonstrated that using the range-separated Coulomb integrals the magnitude of the self-energy in the
Green's function method can be modified as a function of the range separation parameter $\lambda$. These
results demonstrate that srSVWN5---lrGF2 functional can be useful for self-energy embedding calculations as 
well as for Green's function-based calculations of extended systems since for certain values of the parameter
$\lambda$ the decay of self-energy elements is fast and can contribute to an additional sparsity of the
problem. Consequently, a fewer number of self-energy elements need to be evaluated resulting in an overall
reduction of the computational cost. 


Finally, we believe that there are several directions in which short-range DFT with long-range Green's
functions hybrid functionals can be further developed.
In its current implementation the local density functional describes not only the short-range interactions 
but also
the coupling region between short-range and long-range correlations~\cite{Toulouse:ijqc/100/1047}. It has been
shown~\cite{Cornaton:pra/88/022516} that when the coupling region is treated by many-body methods instead of
density functionals , then such a calculation 
results in a further improvement of functional properties.
Therefore, the
development of such range-separated double-hybrid functionals~\cite{Toulouse:tac/114/305} based on 
long-range Green's function methods may be worth pursuing.

Another interesting direction for  the functional proposed in this work is the study of metallic systems or systems with small band gaps. 
Green's function expansions that do not include the infinite sum of bubble diagrams such as a 
M\o ller--Plesset Green's function are experiencing divergences for metallic systems.
These divergences can be
efficiently eliminated by screening of the electron-electron interactions provided by e. g. 
the error function. Therefore, functionals employing a range separation similar to the one presented here, 
may also be applied to periodic calculations of metallic systems in order to avoid a divergent behavior.
Furthermore, several other choices than GF2 such as GW or FLEX   are possible as long-range Green's 
function methods.  
On the density functional side, it is worth investigating short-range semilocal density functionals 
 within the range separation framework. 

\section{acknowledgement}

A.A.K., and D.Z. acknowledge support from 
the U.S. Department of Energy (DOE) grant 
No. ER16391.
A.A.K. was also supported by the University of
Michigan Rackham Predoctoral Fellowship.
A.A.K. is grateful to Dr. Alexander
Rusakov for multiple useful discussions. 



%

\end{document}